\documentstyle[aps,prl,twocolumn]{revtex}
\input epsf 
\draft
\begin{document}

\title{Formation and Pinch-off of Viscous Droplets \\ in the Absence
       of Surface Tension: an Exact Result} 

\author {Mark Mineev-Weinstein, Gary D. Doolen, John E. Pearson}
       \address{Theoretical Division, MS-B213, LANL, Los Alamos, NM
       87545, USA} 
\author {Silvina Ponce Dawson}
\address {Departamento de F\'\i sica, Facultad de
       Ciencias Exactas y Naturales,\\ U.B.A., Ciudad Universitaria,
       Pabell\'on I, (1428) Buenos Aires, Argentina } 

\maketitle
\begin{abstract}
Within a class of exact time-dependent non-singular $N$-logarithmic solutions
[Mineev-Weinstein and Dawson, Phys.~Rev. {\bf E 50}, R24 (1994); 
Dawson and Mineev-Weinstein, 
Phys.~Rev. {\bf E  57}, 3063 (1998)], we have found 
solutions which describe the development and pinching off of viscous droplets 
in the Hele-Shaw cell in the absence of surface tension. 
\end{abstract}

The process of viscous fingering in the Hele-Shaw cell, in which a
non-viscous fluid pushes a viscous one, has attracted the attention of
many physicists and mathematicians since 1958 [1]. The pressure field
in the viscous fluid is harmonic because of D'Arcy's law, $ \mbox{\bf
v} = - \nabla p $, incompressibility, and continuity,
$\mbox{div}\,{\bf v} = 0$, where {\bf v} is fluid velocity. Because
pressure is constant along the moving interface between the
non-viscous fluid and the viscous fluid in the absence of surface
tension, the time-dependent conformal technique was found to be of
great help. The integro-differential equation
\begin{equation}
{\rm Im}(\bar z_t z_{\phi}) = 1\,\,
\label{eq:lge}
\end{equation}
for the moving complex interface $z(t, \phi)$ was derived [2]. Here
the bar denotes complex conjugation, $z_t$ and $z_{\phi}$ are partial
derivatives, and the map $z(t, \phi)$ is conformal for Im $\phi \leq
0$. This equation which we call the Laplacian growth equation (LGE)
has many remarkable properties \cite{richardson}-\cite{silvina}. In
this paper, we make use of a broad class of exact time-dependent
non-singular $N$-logarithmic solutions in the channel geometry (where
the interface moves between the two parallel walls of the Hele-Shaw
cell \cite{mark}) and in the radial geometry (where the moving contour
is closed \cite{silvina})). The equation (\ref{eq:lge}) is valid for
both of these geometries provided the flux of fluid is constant.  In
the channel geometry, the $N$-logarithmic solution has the form
\cite{mark}
\begin{equation}
z(t, \phi) = \tau(t) + i\phi + \sum_{k=1}^{N}\alpha_k \log(1 - a_k(t)
e^{-i\phi}) \,\label{eq:channel},
\end{equation}
where $\alpha_k = {\rm const}$, and $|a_k| < 1$.
The time dependence of $a_k(t)$ and $\tau(t)$ is given by
\begin{eqnarray}
\left\{
\begin{array}{l} 
\beta_k = z(t, i\,\log \bar a_k) \\ = \tau - \log \bar a_k +
\sum_{l=1}^{N}\alpha_l\log(1-\bar a_k a_l) = const, \\ t + C =
\Big(1-\frac{1}{2} \sum_{l=1}^N \alpha_l \Big) \tau +
\frac{1}{2}\sum_{l=1}^{N} \alpha_l \log(a_l) \,,
\label{eq:time_channel}
\end{array}
\right.
\end{eqnarray}
where $k = 1, 2, ..., N$ and $C$ are constants in time. The Hele-Shaw
cell width is chosen to be 2$\pi$. Hence $z(2\pi) - z(0) = 2\pi
i$. This is for periodic boundary condition, while for the no-flux
boundary conditions the same formulas
(\ref{eq:channel}-\ref{eq:time_channel}) are valid, but the sums in
(\ref{eq:channel}-\ref{eq:time_channel}) contain both terms with
$\alpha_k$ and $a_k$ and with $\bar \alpha_k$ and $\bar a_k$ unless
both $\alpha_k$ and $a_k$ are real.

The $N$-logarithmic solution in radial geometry is
\begin{equation}
z(t, \phi) = r(t) e^{i\phi}+\sum_{k=1}^{N}\alpha_k
\log(\frac{e^{i\phi}}{a_k(t)} - 1) \,\label{eq:radial},
\end{equation}
where $\sum_{k=1}^{N}\alpha_k = 0$, $\alpha_k = {\rm const}$, and $|a_k| < 1$. 

Time dependence of $a_k(t)$ and $r(t)$ is given by
\begin{eqnarray}
\left\{
\begin{array}{l} 
\beta_k = z(t, i\,\log \bar a_k) \\
= \frac{r}{\bar a_k}+\sum_{l=1}^{N}\alpha_l\log(\frac{1}{\bar a_k a_l} - 1) \\
= \rm{const}\,\nonumber  \\ 
2t + C =  r^2  - r \sum_{k=1}^{N} \frac{\alpha_k}{a_k} \,.
\end{array}
\right.
\end{eqnarray}
For a broad range of \{$\alpha_k$'s\} (which we do not specify here) the 
solutions (\ref{eq:channel}) are free of finite-time singularities (cusps)
\cite{mark}-\cite{silvina}.

In this paper we present a subclass of solutions (\ref{eq:channel})
which describes the change of topology of the moving interface which
occurs in the process of droplet formation of the viscous fluid. The
moving interface continues to evolve in accordance with the equation
(\ref{eq:lge}). We have found the analytic continuation from the
domain immediately before the topological breakdown occurred to the
domain immediately after the droplet formation. This continuation
allows us to study the interface dynamics which includes an arbitrary
number of topological changes of this kind in the absence of surface
tension. We believe that inclusion of surface tension in these studies
will shed light on which topological changes are physically allowed
and which are forbidden. The goal of this article is to report a new
class of solutions which describes droplet formation in the absence of
surface tension.

Let us consider the following 3-logarithmic solution of the LGE (\ref{eq:lge}) 
in channel geometry:
\begin{equation}
z(t, \phi) = \tau(t) + i\phi +\alpha \log \frac{e^{i\phi} - a(t)}{
e^{i\phi} - b(t)}\,\label{eq:three_log},
\end{equation}
where $\alpha > 0$, $0 < b(0) < a(0) <1$, and the third logarithmic
singularity, corresponding to the term $i\phi$, is zero. It follows
from (\ref{eq:time_channel}) that the time evolution of $a(t)$ and
$b(t)$ obeys the equations
\begin{eqnarray}
\left\{
\begin{array}{l} 
\beta_1 = t - \frac{\alpha}{2} \log \frac{a}{b} + \alpha \log \frac{1 - a^2}{1 - ab}-\log a\\
\beta_2 = t - \frac{\alpha}{2} \log \frac{a}{b} + \alpha \log \frac{1 - ab}{1 - b^2}-\log b\\ 
\label{eq:beta_n_t}
\end{array}
\right.
\end{eqnarray}
Both $a(t)$ and $b(t)$ reach the unit circle simultaneously at the moment, 
$t^*$, which is easy to calculate:

Since $a$ and $b$ are close to 1, then deviations of $a$ and $b$ from 1, 
defined as $p=1-a$ and $q=1-b$, are very small, that is 
$0 < p = 1 - a < q = 1 - b << 1$.
Thus the last system in the leading order is
\begin{eqnarray}
\left\{
\begin{array}{l} 
\beta_1 = t^* + \alpha \log c - \alpha \log \frac{c + c^{-1}}{2}\\
\beta_2 = t^* + \alpha \log c + \alpha \log \frac{c + c^{-1}}{2}\\
\label{eq:beta_n_t-star}
\end{array}
\right.
\end{eqnarray}
where $c^2 = \lim_{t\rightarrow t^*} p/q$ is a finite constant (assuming that 
$t^*$ is finite). The solution of (\ref{eq:beta_n_t-star}) is
\begin{eqnarray} 
&&c   = A - \sqrt{A^2 - 1}\\
\label{eq:c} 
&&t^* = \beta_1 - \alpha \log (1 - \sqrt{1 - e^{(\beta_1 - \beta_2)/\alpha}}, 
\end{eqnarray} 
where $A = e^{(\beta_2 - \beta_1)/2\alpha}$.

It is also easy to calculate the behavior of $a(t)$ and $b(t)$ immediately 
prior to the moment $t^*$:

Defining $\delta t = t^* - t$ and expanding $a(t)$ and $b(t)$ near $t^*$ 
\begin{eqnarray} 
\left\{
\begin{array}{l} 
a = 1 - p_0 \delta t - p_1 (\delta t)^2\\
b = 1 - q_0 \delta t - q_1 (\delta t)^2,
\end{array}
\right.
\end{eqnarray}
(clearly then, that $p_0/q_0=c^2$), 
we rewrite (\ref{eq:beta_n_t}) keeping linear terms with respect to $\delta t$:
\begin{eqnarray}
\left\{
\begin{array}{l} 
1 = \alpha (\frac{p_1}{p_0} - \frac{p_1 + p_0}{q_1 + q_0}) + p_0 (\frac{
\alpha}{2}\, \frac{q_0 - p_0}{q_0 + p_0} + 1)\\ 
1 = \alpha (\frac{p_1 + p_0}{q_1 + q_0} - \frac{q_1}{q_0}) + q_0 (\frac{
\alpha}{2}\, \frac{q_0 - p_0}{q_0 + p_0} + 1) 
\end{array}
\right.\end{eqnarray}
Keeping in mind that $c^2 = p_0/q_0$ (see above) we find from this last system 
that
\begin{eqnarray}
\left\{
\begin{array}{l}
p_0 = \frac{c^2}{(1 + \alpha/2) + c^2(1 - \alpha/2)}\\
q_0 = \frac{1}{(1 + \alpha/2) + c^2(1 - \alpha/2)},
\end{array}
\right.\end{eqnarray}
where $c$ is given by (\ref{eq:c}).
\begin{figure}
\epsfxsize=2.25truein
\epsfysize=2.25truein
\hskip .5truein
\epsfbox{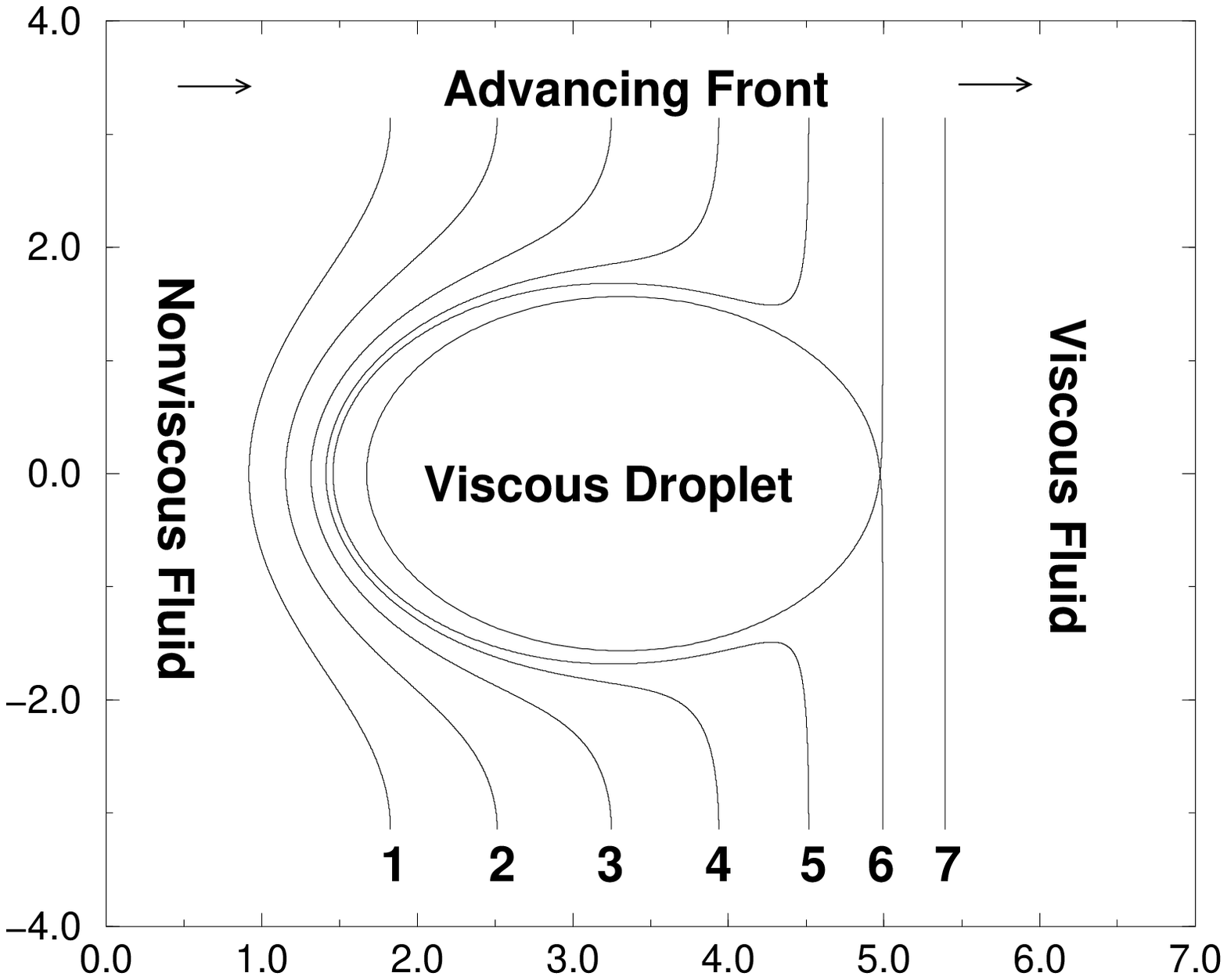}
\caption{
The front, moving to the right, forms a viscous droplet, and then leaves it
behind. The numbers 1-7 denote 7 distinct snapshots of the advancing front.
}

\label{fig:droplet}
\end{figure}

It is interesting to see how the advancing front moves in this case.
Figure~\ref{fig:droplet} shows that the initially almost planar
interface ($b(0) << 1$) advances until a groove of width $\pi \alpha$
is created at the stagnation point with complex coordinate $\beta_1 -
\alpha \log 2$. Then later this groove becomes thinner and thinner
until at the moment $t^*$ it disappears completely. Since both $a(t)$
and $b(t)$ reach the unit circle simultaneously, (\ref{eq:three_log})
contains a singular term $\rm{log} [(e^{i\phi} - 1)/(e^{i\phi} - 1)]$
which has a removable singularity at 1.  This term equals zero {\it
exactly at this moment of time}. If we formally consider $t > t^*$ in
(6) we see that both $a(t)$ and $b(t)$ continue their evolution
outside the unit circle. This is clearly an unphysical part of the
solution (\ref{eq:three_log}), which violates the framework of
validity of this solution, which demands that the exterior of the unit
circle should be free of singularities except at infinity. We also
note that the range of the variable $\phi$ parametrizing the moving
curve, corresponding to the groove of viscous liquid which is left
behind the advancing front, becomes negligibly small when the ``neck''
of this groove becomes very thin.  This allows us to make the
analytical continuation from the moment $t^*_-$ immediately prior to
the disappearance of the neck and pinching off the viscous droplet
behind the moving interface to the moment $t^*_+$ that is immediately
after this event occurs. Once both singularities $a(t)$ and $b(t)$
reach the unit circle at $t = t^*$, the last term in
(\ref{eq:three_log}) vanishes and the subsequent evolution ignores
this term. The interface in this case becomes a planar front described
by the equation
\begin{equation}
z(t, \phi) = \tau(t) + i\phi\,,
\end{equation}
where $\tau = t$ in this case.  In the physical plane, this analytical
continuation at the moment $t = t^*$ shows that the viscous droplet,
which was formed from the original groove at the stagnation point
$\beta_1 - \alpha \log 2$, pinches off from the advancing interface
and is left behind the moving front. The moving front for $t > t^*$ is
described by the same solution but without the pair of logarithms.
This process can be seen in Figure~\ref{fig:droplet}. The shape of the
droplet does not change after $t = t^*$. This is because of two
limitations of the present theory: both the viscosity of the less
viscous liquid and the surface tension were chosen to be zero. This
leads the pressure along the droplet boundary to be a constant, and
then, by virtue of the maximum theorem for harmonic functions, the
pressure should be constant throughout the interior of the
droplet. The pressure gradient acting on the droplet is
zero. Consequently the droplet is at rest in the non-viscous
domain. In the real situation, where neither the second liquid
viscosity nor the surface tension is zero, the viscous droplet will be
dragged by the flow and its shape will continue to be modified.

The shape of the droplet described by (\ref{eq:three_log}) has a
simple form: this is an oval with a horizontal length of the order of
$\beta_2 - \beta_1 + 2\alpha {\rm log} 2$ and the vertical size about
$\pi \alpha$. The droplet's shape can be, however, much more
complex. It can be described by the N-logarithmic solution
(\ref{eq:channel}) such that the $\alpha_k$ corresponding to the
minimal $|a_k|$ is chosen to be $ - \sum'_l \alpha_l$, where the prime
in the sum indicates summation over all $l$ from $1$ to $N$, but
omitting $l=k$.

In \cite{mark} the geometrical interpretation of the constants of
motion $\{\alpha_k\}$ and $\{\beta_k\}$ have been found. Namely,
$\beta_k - \alpha_k \log 2$ is the complex coordinate of the
stagnation point near which the $k^{th}$ groove with parallel walls
originates. And $\pi|\alpha_k|$ and arg($\alpha_k$) are the width and
the angle with respect to the horizontal axes respectively. This
geometrical interpretation appeared to be in the excellent agreement
with known real and numerical experiments. But this interpretation has
a constraint: $\vert {\rm arg}(\alpha_k)\vert < \pi/2$. In this work
we consider $\alpha_k$ which violates the last inequality. For such
$\alpha_k$ the $\beta_k - \alpha_k {\rm log}2$ is the complex
coordinate of the point where the previously formed groove starts to
shrink. Clearly, for the simplest example with two logarithms, where
$\alpha_1 > 0$ and $\alpha_2 < 0$, the width of the groove which was
originally $\pi \alpha_1$ becomes after shrinking $\pi(\alpha_1 +
\alpha_2)$. Of course, this consideration is valid only when $\alpha_1
+ \alpha_2 \geq 0$. In the case of equality the width vanishes
completely, it corresponds to the formation of the droplet described
above.  We do not understand on this stage why and when the terms with
Im($\alpha_k) < 0$ are physically realizable. We hope that future
studies will elucidate this point.

Let us derive the droplet's neck dynamics $w(t)$ prior to pinch off in 
the example (\ref{eq:three_log}) discussed above. The width $w(t)$ of the neck 
is defined as
\begin{equation}
w(t) = 2y(t, \phi^*(t)) = 2{\rm Im}\, z(t, \phi^*(t)
\label{eq:w},
\end{equation}
where $\phi^*$ is a non-zero solution of the equation
\begin{equation}
\partial y / \partial \phi = 0
\label{eq:der},
\end{equation}
as it is clear from (\ref{eq:three_log}) and from
Figure~\ref{fig:droplet}.  Differentiating (\ref{eq:three_log}) and
solving (\ref{eq:der}), having in mind that both $a$ and $b$ are close
to 1, we obtain $\phi^* = \sqrt{\alpha (a-b)}$ in the leading order
with respect to $(a-b)$ which is small. Then from (\ref{eq:w}) and
(\ref{eq:three_log}) it follows that
\begin{equation}
w(t) = 2[\phi^* + \alpha ( \arctan \frac{ \sin \phi^*}{ \cos \phi^* - 
a} - \arctan \frac{ \sin \phi^*}{ \cos \phi^* - b})]
\label{eq:width_sol}.
\end{equation}
Replacing here $\phi^*$ by its value found above we finally obtain
\begin{equation}
w(t) = 4 \sqrt{\alpha (a-b)} = 4 \sqrt{\alpha (1-c^2) q_0 (t^* - t)},
\label{eq:width_final}
\end{equation}
where constants $c$, $q_0$, and $t^*$ are given above.

Here we considered only the simplest configuration
(\ref{eq:three_log}) leading to the formation of a single
droplet. Clearly, there is a broad subclass of solutions
(\ref{eq:channel}) and (\ref{eq:radial}) which describes in a similar
fashion formation and pinching off of an arbitrary number of viscous
droplets with different shapes. The same analytic continuation,
performed above for the simple case $N=3$, holds for this generalized
case. Again, this continuation contains the elimination of the
removable singularity by equating $(e^{i\phi} -1)/(e^{i\phi} -1)$ to 1
at the moment of formation of each droplet, thus decreasing the total
number of logarithms in the description of the moving interface, as
done above for $N=3$.  All these considerations hold for both the
channel geometry and for closed interfaces.

Previous studies on topological changes in the viscous flow in the
Hele-Shaw channel \cite{dupont} considered surface tension as a
significant factor.  Because surface tension is zero in the present
paper and because we believe that there are many different regimes of
pinching off within classes of solutions (\ref{eq:channel}) and
(\ref{eq:radial}), we do not compare our findings with results in
\cite{dupont}. There is also a paper on the change of topology in the
absence of surface tension \cite{etingof}, which states that a change
of topology is inevitable in the interior (suction) problem if the
sink is not located at the center of mass of the viscous domain.  Here
we may say that we know how to describe possible topological changes
in detail. What we do not know is which of the solutions obtained
above are physically realizable and which are not. We also do not know
at this stage what will be the corrections to the dynamics of droplet
formations caused by non-zero surface tension. We hope that future
studies will answer these questions.

S.P.D thanks CONICET and UBA for the support, and other authors
gratefully acknowledge the support of DOE programs during these
studies.

\end{document}